# Usable Speech Assignment for Speaker Identification under Co-Channel Situation

Wajdi Ghezaiel
CEREP-Ecole Sup. des Sciences et Techniques de Tunis, Tunisia

Amel Ben Slimane
Ecole Nationale des Sciences de l'Informatique Mannouba, Tunisia

Ezzedine Ben Braiek
CEREP-Ecole Sup. des Sciences et Techniques de Tunis, Tunisia

## ABSTRACT
Usable speech criteria are proposed to extract minimally corrupted speech for speaker identification (SID) in co-channel speech. In co-channel speech, either speaker can randomly appear as the stronger speaker or the weaker one at a time. Hence, the extracted usable segments are separated in time and need to be organized into speaker streams for SID. In this paper, we focus to organize extracted usable speech segment into a single stream for the same speaker by speaker assignment system. For this, we develop model-based speaker assignment method based on posterior probability and exhaustive search algorithm. Evaluation of this method is performed on TIMIT database. The system is evaluated on co-channel speech and results show a significant improvement.

## 1. INTRODUCTION
Speaker identification (SID) is the task of recognizing the identity of somebody on the observed speech signal. Speaker identification systems consist of short-term spectral feature extractor (front-end) and a pattern matching module (back-end). In traditional SID, only one target speaker exists in the given signal whereas in co-channel SID, the task is to identify the target speakers in one given mixture. Research on co-channel speaker identification has been done for more than one decade [1], yet the problem remains largely unsolved.

Research has been carried to extract one of the speakers from co-channel speech by either enhancing target speech or suppressing interfering speech [1], [2]. Zissman and Seward [3] looked at pitch continuity in co-channel speech and assigned pitch contours to a corresponding talker by polynomial contour fitting when pitch contours from two speakers cross. Their results advice that a method based purely on pitch information is not sufficient. Morgan et al. [2] estimated the dominant pitch and then reconstructed the speech components of both stronger and weaker talker frame by frame using frequency-domain filtering according to the estimated pitch. Afterward, a speaker assignment algorithm using a maximum-likelihood criterion is applied to group recovered signals into two speaker streams, one for the target and the other for the interferer. The assignment algorithm groups the individual frames by inspecting the pitch and spectral continuity for consecutive voiced frames, and comparing the spectral similarity of the onset frame of a voiced segment with recently assigned frames using the divergence measure proposed by Carlson and Clement [4], which is the symmetrized Kullback–Leibler divergence [5].

In automatic speaker recognition, as pointed out in [8], the intelligibility and quality of extracted speech are not important. What the system needs are portions of the speech that contain speaker characteristics unique to an individual speaker, classifiable and long enough for the system to make identification or verification decisions. These portions of speech, or segments, are defined as consecutive frames of speech that are minimally corrupted by interfering speech and are, thus, called usable speech [8]. Several methods based on usable speech measures have been developed and studied under co-channel conditions [9] [10] [11] [12] [13] [14]. These methods show that the speaker identification system could achieve approximately 80% of correct identification when the overall TIR is 20 dB. In co-channel speech, either speaker can randomly appear as the stronger speaker or the weaker one at a time. Hence, the extracted usable segments are separated in time and need to be organized into speaker streams for SID. For this, we propose a speaker assignment system that organizes usable speech segments under co-channel conditions. We extend the probabilistic framework of traditional SID to co-channel speech. In this paper, we propose to organize extracted usable speech, speaker homogeneous segments into streams. Our method employs multi-resolution dyadic wavelet transform MRDWT proposed recently [14] for detection of usable speech. We employ exhaustive search algorithm to maximize the posterior probability in grouping usable speech. Then, usable segments are assigned to two speaker groups, corresponding to the two speakers in the mixture. Finally, speakers are identified using the assigned segments. The next section we describe how to extract usable speech using multi-resolution analysis of dyadic wavelet transform. In section 3, we develop the proposed speaker assignment system. Experiment results and comparisons are given in section 4. Section 5 concludes the paper.

## 2. Multi-Resolution Dyadic wavelet Transform Method (MRDWT) for Usable Speech Detection
Usable frames are characterized by periodicity features. These features should be located in low-frequency band that includes the pitch frequency. Multi-resolution analysis based on DWT [14] is applied iteratively in order to determine the suitable band for periodicity detection In this band periodicity features are not much disturbed by interferer speech in case of usable segments. In case of unusable frames it is not possible to detect periodicity in all lower sub-bands until the pitch band of a male speaker. Pitch band of a male speaker is lower than a female speaker pitch band so that it is imperative to filter until this band. At each scale, autocorrelation is applied to the approximation coefficients in order to detect periodicity. Three maxima are determined from the autocorrelation signal with a peak-picking algorithm using threshold on local maximum amplitude. A difference of autocorrelation lag between the first and second maxima and the autocorrelation lag between the second and third maxima is determined. This difference value notices the presence of pitch information. If this difference is less than the preset threshold periodicity is detected. Figure 1 shows that periodicity is not detected at scale 1 but it is detected at scale 2. Figure 2 corresponds to a usable frame for male-male co-channel speech. In this case, Periodicity is detected only at scale 3.

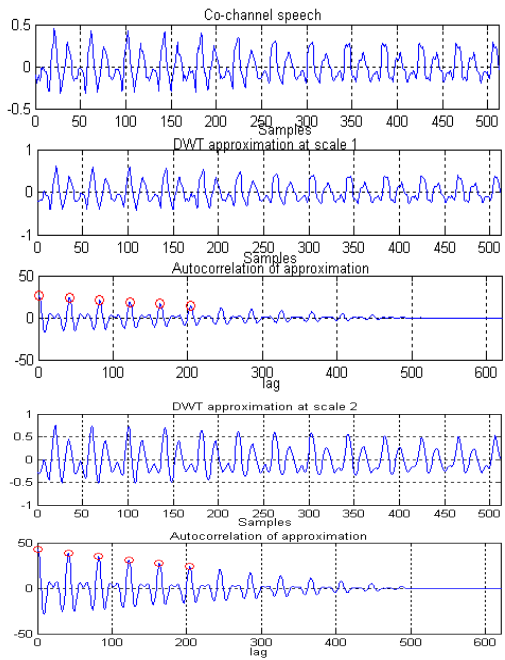

**Fig 1: Analysis of a usable speech frame for female-female co-channel speech up to scale 2.**

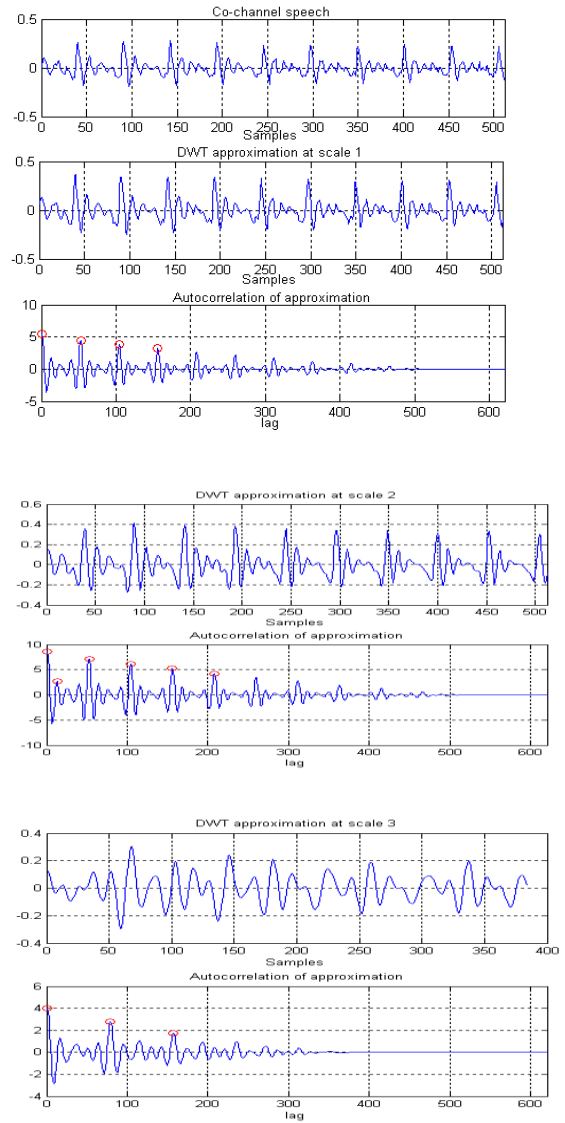

**Fig 2: Analysis of a usable speech frame for male-male co-channel speech up to scale 3.**

Figure 3 presents unusable frame speech. This frame is classified us unusable because periodicity was not detected at all scales.

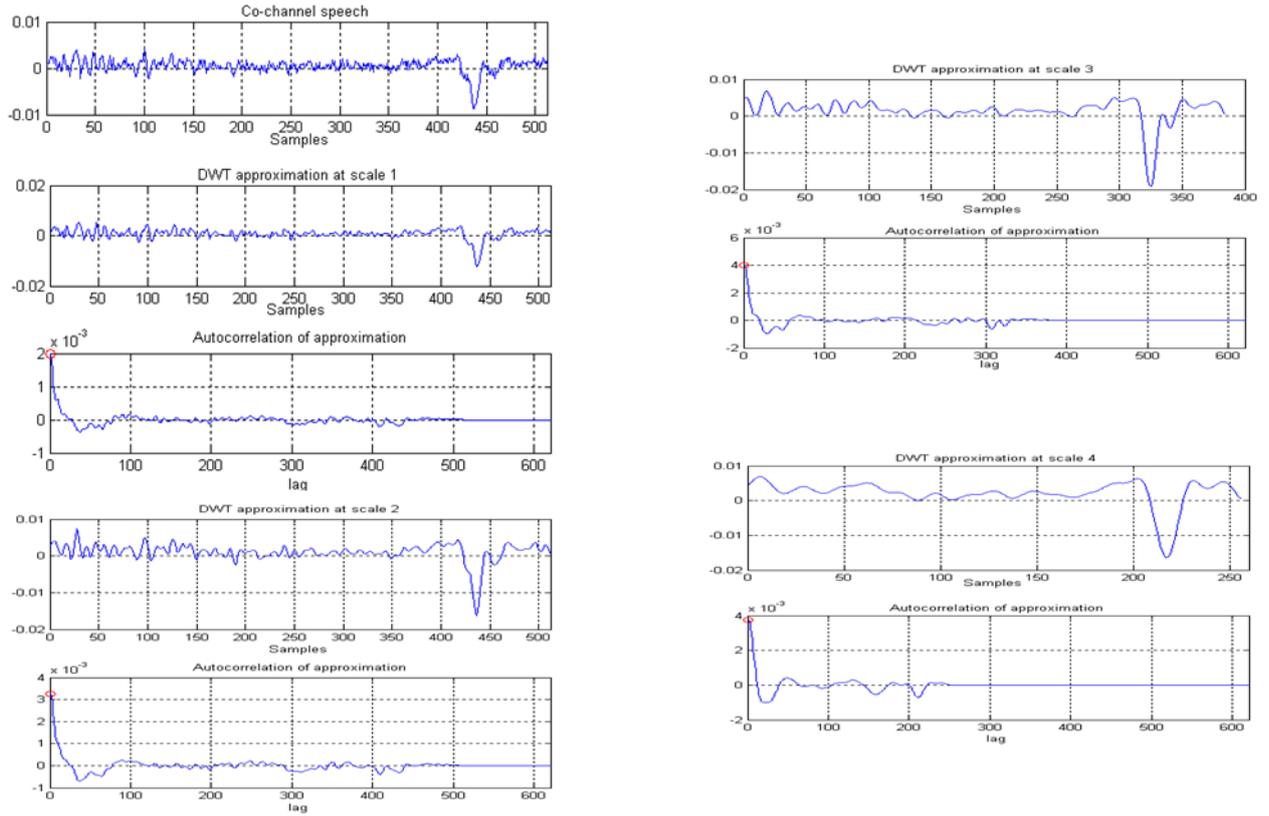

**Fig 3: Analysis of a unusable speech frame up to scale 4.**

## 3. Usable Speech Assignment

In speaker identification system, discrimination between speakers is based on posterior probability. The goal is to find the speaker model reference in the set of speaker models $\Lambda = \{\lambda_1, \lambda_2, \ldots, \lambda_k\}$, that maximizes the posterior probability for an observation sequence $O = \{o_1, o_2, \ldots, o_M\}$. Cepstral features, such as mel-frequency cepstral coefficients (MFCCs), are used as observations for speech signals. The SID decision rule is  $\lambda_I' = \mathrm{argmax}\, P(\lambda\, O)\, ; \lambda \in \Lambda$   (1)

Applying the Bayesian rule, we have $\lambda_I' = \mathrm{argmax}\, \frac{P(O\,\lambda)\, P(\lambda)}{P(O)}$   (2)

Typically, prior probabilities of speakers are equal, and the maximization over $\lambda$ is not affected by $P(O)$. Hence, $P(\lambda)$ and $P(O)$ can be dropped. Using speaker models and assuming independence between observations at different times, (2) can be re-written as: $\lambda_I' = \mathrm{argmax}\, \sum_{m=1}^{M} \log p(o_m\, \lambda)$  (3) after taking the log operation. Here, m indexes observations. $p(o|\lambda)$ is the standard Gaussian mixture model estimated from training speech of specific talkers using the EM algorithm [15].

The problem in co-channel attempts to find two speaker models that maximize the posterior probability for the observations. For a co-channel mixture, our usable speech extraction method extracts N consecutive speech segments, $X = \{S_1, S_2, \ldots, S_i, \ldots, S_N\}$. Given X, (1) can be modified as follows:

$\lambda_I', \lambda_{II}' = \mathrm{argmax}\, P(\lambda_I, \lambda_{II}|X)\, ; \lambda_I, \lambda_{II} \in \Lambda$   (4)

which is to provide a pair of speaker models, $\lambda_I'$ and $\lambda_{II}'$, from the speaker set $\Lambda$ that maximize the posterior probability given usable speech segments. Usable segments must be organized into two speaker streams because in co-channel speech one speaker can dominate in some portions and be dominated in other portions. For example, a possible segment assignment may look like $S_1^0, S_2^1, \ldots, S_i^0, \ldots, S_N^1$, where superscripts, 0 and 1, do not represent the speaker identities but only indicate that the segments marked with the same label are from the same speaker. Therefore, the objective of speaker assignment is tracking a pair of speaker models, $\lambda_I'$ and $\lambda_{II}'$, together with a segment assignment, $y'$, that maximize the posterior probability:

$\lambda_I', \lambda_{II}', y' = \mathrm{argmax}\, P(\lambda_I, \lambda_{II}, y\, X)\, ; \lambda_I, \lambda_{II} \in \Lambda, y \in Y$   (5)

Where Y is the assignment space, which includes all possible assignments (labeling) of the segments.

The posterior probability in (5) can be rewritten as

$P(\lambda_I, \lambda_{II}, y, X) = \mathrm{argmax}\, \frac{P(\lambda_I, \lambda_{II}, y, X)}{P(X)}$

$= P(X\, y, \lambda_I, \lambda_{II}) P(y/\lambda_I, \lambda_{II}) \frac{P(\lambda_I, \lambda_{II})}{P(X)}$   (6)

Since the assignment is independent of specific models, $P(y/\lambda_I, \lambda_{II})$ becomes $P(y)$, which, without prior knowledge on segment assignment, we assume to be uniformly distributed. Assuming the independence of speaker models and using the same assumption from traditional SID that prior probabilities of speaker models are the same, we insert (6) into (5) and remove the constant terms. The objective then becomes finding two speakers and an assignment that have the maximum probability of assigned usable speech segments given the corresponding speaker models as follows:

$$\lambda'_I, \lambda'_{II}, y' = \text{argmax } P(X\ y, \lambda_I, \lambda_{II}) \quad (7)$$

Note the conditional probability is essentially the joint SID score of assigned segments. Given y, the labeling, we denote $X^0$ as the subset of usable speech segments labeled 0, and $X^1$ the subset labeled 1. Since $X^0$ and $X^1$ are complementary, the probability term in (7) can be written as follows:

$$P(X\ /\ y, \lambda_I, \lambda_{II}) = P(X^0, X^1 | \lambda_I, \lambda_{II}) \quad (8)$$

The y term is dropped from the above equation because the two subsets already incorporate the labeling information. Assuming that any two segments, $S_i$ and $S_j$, are independent of each other given the speaker models and that segments with different labels are produced by different speakers, the conditional probability in (8) can be written as

$$P(X^0, X^1 | \lambda_I, \lambda_{II}) = P(X^0\ \lambda_I, \lambda_{II})\ P(X^1\ \lambda_I, \lambda_{II})$$
$$= \prod_{S_i \in X^0} P(S_i\ \lambda_I) \prod_{S_j \in X^1} P(S_j | \lambda_{II}) \quad (9)$$

The probability of having a segment S from a pre-trained speaker model $\lambda$ is the product of likelihoods of that speaker model generating each individual observation x of the segment, assuming the observations are independent of each other. In other words $P(S\ \lambda) = \prod_{x \in S} p(x | \lambda) \quad (10)$

The goal is to find two speakers and one assignment that yield the maximal probability using (10). Given the extracted usable speech segments and individual speaker models trained from clean speech, the problem is to search two speakers in space $\Lambda$ and Y that maximize probability in (10). The brute-force way to find the maximum is exhaustive search. We employ exhaustive search algorithm to find correspondent speaker. In implementation, the real computation time is longer. It can be further reduced by storing all the likelihood scores of a segment given a model in the memory as a table and looking up a score from the table when needed.

## 4. Experiment and result

We employ the evaluation data from the TIMIT speech corpus. The speaker set is composed of 38 speakers from the "DR1" dialect region, 14 of which are female and the rest are male. Each speaker has 10 utterance files. For each speaker, 5 out of 10 files are used for training and the remaining 5 files are used to create co-channel mixtures for testing. For each speaker deemed as the target speaker, 1 out of 5 test files is randomly selected and mixed with randomly selected files of every other speaker, which are regarded as interfering utterances. For each pair the TIR is calculated as the energy ratio of the target speech over the interference speech. Speech signals are scaled to create the mixtures at different TIRs: -20 dB, -10 dB, -5 dB, 0 dB, 5 dB, 10 dB and 20 dB. For example, 0 dB TIR means that the overall energy of target is equal to that of interference. Three different sets of co-channel speech are considered: male-male, female-female, and male-female.

### 4.1 Speaker Assignment Evaluation

Here, we evaluate the performance of our speaker assignment system. For this evaluation, we only consider co-channel mixtures with overall TIR equal to 0 dB to simulate real co-channel situations. Table I present evaluation results. It includes results from alternative methods for speaker assignment. In table I, the baseline rate of correct assignment corresponding to random labeling of each usable frame is 50.0%. The 2nd row shows that ideal assignment by prior pitch achieves 94.1% correct rate. Exhaustive search achieves 86.4% correct assignment rate. It reflects the effectiveness of using speaker characteristics and exhaustive search algorithm for speaker assignment.

**Table.1 Evaluation results of speaker assignment methods**

| Method | Speaker assignment % |
| --- | --- |
| Random assignment | 50.00 |
| Ideal assignment by prior pitch | 94.1 |
| Exhaustive search | 86.4 |

### 4.2 Speaker Identification Evaluation

We employ 16 MFCCs coefficients as speaker features. Speakers are modeled using 16-mixture GMMs, which are trained using the EM algorithm [15] from the training samples. As a baseline, a conventional SID system is applied to the co-channel speech to recognize the target speaker. Figure 4 gives the target speaker recognition accuracy. The accuracy degrades sharply when TIR decreases because the target speech is increasingly corrupted. The first observation from the figure is that, under co-channel situations, usable speech extraction and speaker assignment system improves significantly SID performances. Secondly, the improvements are consistent across all TIR levels. Performance improvement increases at higher TIRs because the target speaker dominates the mixture. However, target speaker is dominated by interference at lower TIRs, resulting in lower performance after usable speech extraction.

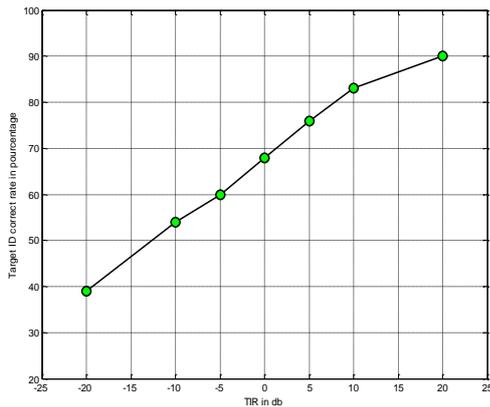

**Fig 4: Target SID correct rate**

## 5. Conclusion

In this paper, we have proposed a speaker assignment system to organize extracted usable speech. We have employed our new proposed method for usable speech detection based on multi-resolution analysis using dyadic wavelet transform MRDWT. Our usable speech extraction method produces segments useful for co-channel SID across various TIR conditions. Evaluation of the speaker assignment system gives significant results. Application of usable speech and speaker assignment improve the speaker identification performance in co-channel conditions.